\documentstyle[12pt,a4,epsfig,a4wide]{article} 
\def\Jo#1#2#3#4{{#1} {\bf #2}, #3 (#4)}

\def\NPB{{Nucl. Phys.} {\bf B}}
\def\NPBP{{Nucl. Phys.} {\bf B} (Proc. Suppl.)}

\def\PLB{{Phys. Lett.}  {\bf B}}
\def\PRL{Phys. Rev. Lett.}
\def\PRD{{Phys. Rev.} {\bf D}}

\def\PRP{{Phys. Rep. }}

\def\ZPC{{Z. Phys.} C}

\def\PNPP{Prog. Nucl. Part. Phys.}

\def\IJMP{Int. J. Mod. Phys. {\bf A}}

\def\ra{\rightarrow}
\def\be{\begin{equation}}
\def\ee{\end{equation}}
\def\gs{\mathrel{
   \rlap{\raise 0.511ex \hbox{$>$}}{\lower 0.511ex \hbox{$\sim$}}}}
\def\ls{\mathrel{
   \rlap{\raise 0.511ex \hbox{$<$}}{\lower 0.511ex \hbox{$\sim$}}}}

\newcommand{\onbb}{neutrinoless double beta decay}

\newcommand{\ba}{\begin{array}{c}}
\newcommand{\baz}{\begin{array}{cc}}
\newcommand{\bad}{\begin{array}{ccc}}
\newcommand{\bav}{\begin{array}{cccc}}
\newcommand{\bea}{\begin{equation} \begin{array}{c}}
\newcommand{\eea}{ \end{array} \end{equation}}
\newcommand{\ea}{\end{array}}
\newcommand{\D}{\displaystyle}

\newcommand{\meff}{\mbox{$\langle m \rangle$ }}

\newcommand{\dma}{\mbox{$\Delta m^2_A$}}
\newcommand{\dms}{\mbox{$\Delta m^2_\odot$}}

\begin{document}

\title{  
\hfill { \bf {\small hep-ph/0105175}}\\ 
\hfill { \bf {\small DO--TH 01/06}}\\ \vskip 1cm 
\bf A simple connection between neutrino oscillation and leptogenesis}
\author{Anjan S.~Joshipura$^a$\thanks{E--mail: \tt anjan@prl.ernet.in}, 
$\;$ Emmanuel A.~Paschos$^b$\thanks{E--mail: \tt paschos@physik.uni-dortmund.de}, $\;$ Werner Rodejohann$^b$\thanks{E--mail: \tt 
rodejoha@xena.physik.uni-dortmund.de}\\ \\
{ \normalsize \it $^a$Theoretical Physics Group,} 
{ \normalsize \it Physical Research Laboratory,}\\ 
{\normalsize \it Navarangapura, Ahmedabad, 380 009, India}\\ \\
{\normalsize \it $^b$Institut f\"ur Theoretische Physik, 
Universit\"at Dortmund,}\\
{\normalsize \it Otto--Hahn--Str.4, 44221 Dortmund, Germany}}
\date{}
\maketitle
\thispagestyle{empty}
\begin{abstract}
The usual see--saw formula is modified
by the presence of  two Higgs triplets in left--right symmetric theories. 
The contribution from the left--handed Higgs triplet 
to the see--saw formula can dominate over the conventional one when 
the neutrino Dirac mass matrix is identified with the 
charged lepton or down quark mass matrix.  
In this case an analytic calculation of the lepton asymmetry, 
generated by the decay of the lightest right--handed Majorana neutrino, 
is possible. 
For typical parameters, the out--of--equilibrium condition 
for the decay is automatically 
fulfilled. The baryon asymmetry has the correct order of magnitude, 
as long as the lightest mass eigenstate is not much lighter then 
$10^{-6}$ to $10^{-8}$ eV, depending on the solution of the solar 
neutrino problem. 
A sizable signal in neutrinoless double beta decay can be expected, 
as long as the smallest mass eigenstate is not much lighter than 
$10^{-3}$ eV and the Dirac mass matrix is identified with the 
charged lepton mass matrix.   
\end{abstract}

\newpage

\section{\label{sec:intro}Introduction}
The impressive evidence for non--vanishing neutrino masses \cite{nurev} 
opens the possibility to study a number of new physics problems on 
a broader phenomenological basis. For the first time 
non Standard Model physics is seen. 
For example, the observed baryon asymmetry of the universe cannot 
be explained by conventional Standard Model physics \cite{SMnot}. Since 
the leptogenesis mechanism proposed by Fukugita and Yanagida \cite{first} 
requires very massive right--handed neutrinos, it is a  fruitful 
task to try to connect the properties of the known light neutrinos 
with their heavy right--handed counterparts. For this, certain 
assumptions concerning the high--energy part of the theory 
have to be made. 
Several recent articles dealt with this subject 
\cite{lola,laza1,others,kang,goldberg,nezri,faltra1,faltra2,nielsen}. 
In \cite{eapjos} we argued that the left--handed Higgs triplet 
--- which was absent in the previous papers --- gives the most important 
contribution to the neutrino mass 
matrix (see also Refs.\ \cite{chun}). 
In addition to the triplet contribution there is a correction from the 
usual see--saw \cite{seesaw} term $m_D \, M_R^{-1} \, m_D^T$.
The large top quark mass makes this see--saw contribution comparable to 
the triplet contribution 
if $m_D$ is identified with the up quark mass matrix. 
This is no longer the case if
the mass scales in $m_D$ are determined by the charged 
lepton or down quark masses.
The neutrino mass matrix is then described mainly by the Higgs 
triplet contribution. 
This has the remarkable consequence that 
the mixing matrix, which diagonalizes the light left--handed 
neutrino mass matrix and is measured in oscillation experiments, is 
{\it identical} to the one which diagonalizes the heavy right--handed 
mass matrix, which governs the lepton asymmetry. In addition, 
the masses of the heavy right--handed Majorana neutrinos are 
proportional to the masses of the light left--handed ones. 
Apart from this aesthetically attractive property, the 
scenario allows to derive very simple analytical 
results on the baryon asymmetry in terms of the measured oscillation 
parameters and the yet unknown phases in the mixing matrix. 
We find that 
with natural choice of parameters the observed  
value of \cite{citeYB}
\be \label{eq:YBexp}
Y_B \simeq (0.1 \ldots 1) \cdot 10^{-10} , 
\ee
is predicted. Within our scenario, it is possible to 
obtain a limit on the lightest mass eigenstate $m_1$, 
which is $10^{-6}$ ($10^{-7} ,\, 10^{-8}$) 
eV for the small mixing (large mixing, vacuum) solution of the 
solar neutrino problem. 
Furthermore, the 
out--of--equilibrium condition $\Gamma_1 \ls H(M_1)$, where $\Gamma_1$ 
is the decay width of the lightest heavy Majorana and $H(M_1)$ 
the Hubble constant at the time of the decay, is also fulfilled. 
Finally, we make the connection to neutrinoless double beta decay 
and find that 
a sizable signal can be expected, 
as long as the smallest mass eigenstate is not much lighter than 
$10^{-3}$ eV and the Dirac mass matrix is identified with the 
charged lepton mass matrix.\\

The paper is organized as follows: In Section \ref{sec:oscleplr} we 
summarize how leptogenesis and neutrino oscillations are connected 
in left--right symmetric theories. This section 
summarizes the framework described in our earlier paper \cite{eapjos}. 
In Section \ref{sec:estYB} we perform an estimate of the baryon asymmetry, 
which explains almost all basic features found in the detailed 
numerical analysis, presented in Section \ref{sec:res}. 
The interesting connection to neutrinoless double beta decay 
is drawn in Section \ref{sec:0vbb}. 
Our conclusions are presented in Section \ref{sec:concl}.

\section{\label{sec:oscleplr}Neutrino Oscillation and Leptogenesis in 
left--right symmetric theories}
The light and heavy neutrino masses are obtained by diagonalizing 
\be \label{eq:mogem}
\left( \baz m_L           & \tilde{m}_D \\[0.3cm]
            \tilde{m}_D^T & M_R 
 \ea \right)  , 
\ee
where $m_L$ ($M_R$) is a left--handed (right--handed) Majorana 
and $\tilde{m}_D$ a Dirac mass matrix. 
This yields  
\be \label{eq:mnu}
m_\nu = m_L - \tilde{m}_D \, M_R^{-1} \, \tilde{m}_D^T . 
\ee
The contribution of $m_L$ is often neglected, even though it might 
play an important role in explaining the oscillation data \cite{moh}. 
This matrix is further diagonalized by 
\be \label{eq:uldef}
U_L^T \, m_\nu \, U_L = {\rm diag} (m_1,m_2,m_3) , 
\ee
where $m_i$ are the light neutrino masses. The entries of $U_L$ are 
then measured in oscillation experiments, see Section \ref{sec:mnu}. 
The symmetric matrix $M_R$ also appears in the Lagrangian 
\be \label{eq:convss}
-\mbox{$\cal{L}$}_Y 
= \overline{l_{iL}} \, \frac{\D \Phi}{\D \langle \Phi \rangle} \, 
\tilde{m}_{D ij} \, N'_{Rj}
+ \frac{\D 1}{\D 2}\overline{N'^c_{Ri}} \, M_{R ij} \, N'_{Rj} + \; \rm h.c.
\ee
with $l_{iL}$ the leptonic doublet and $\langle \Phi \rangle$ the 
vacuum expectation value (vev) of the Higgs doublet $\Phi$.  
Diagonalizing $M_R$ brings us to the physical basis 
\be \label{eq:urdef}
U_R^T \, M_R \, U_R = {\rm diag}(M_1, M_2, M_3) 
\ee
and defines the physical states 
\be
N_R = U_R^\dagger \, N'_R . 
\ee
In the new basis the Dirac mass matrix appearing in the first part 
of the Lagrangian Eq.\ (\ref{eq:convss}) also changes to
\be \label{eq:mdur}
m_D = \tilde{m}_D \, U_R  . 
\ee
It is the rotated Dirac mass matrix which determines the lepton asymmetry.  
The asymmetry is caused by the interference of tree level with one--loop 
corrections for the decays of the lightest Majorana, $N_1 \ra \Phi \, l^c$ 
and $N_1 \ra \Phi^\dagger \, l$: 
\be \label{eq:eps}
\varepsilon = \frac{\D \Gamma (N_1 \ra \Phi \, l^c) - 
\Gamma (N_1 \ra \Phi^\dagger \, l)}{\D \Gamma (N_1 \ra \Phi \, l^c) +  
\Gamma (N_1 \ra \Phi^\dagger \, l)}
= \frac{\D 1}{\D 8 \, \pi \, v^2} \frac{\D 1}{(m_D^\dagger m_D)_{11}} 
\sum_{j=2,3} {\rm Im} (m_D^\dagger m_D)^2_{1j} \, f(M_j^2/M_1^2) 
\ee
Here, $v \simeq 174$ GeV is the weak scale and the function $f$ includes 
terms from vertex \cite{first,luty} 
and self--energy \cite{leptogenesis} contributions: 
\be \label{eq:fapprox}
f(x) = \sqrt{x} \left(1 + \frac{1}{1 - x} - 
(1 + x) \, \ln \left(\frac{1 + x}{x}\right) \right)
\simeq - \frac{3}{2 \, \sqrt{x}}  . 
\ee
The approximation holds for $x \gg 1$. 
To calculate the asymmetry 
$\varepsilon$, we need the Dirac mass matrix $\tilde{m}_D$ and the 
matrix $U_R$ in order to obtain $m_D$ in 
Eq.\ (\ref{eq:mdur}). 
In general, after introducing $\tilde{m}_D$ in the 
Lagrangian, one may obtain a very different $m_D$ after the rotation 
with $U_R$.\\ 

In our approach, the left--right symmetry \cite{LR} plays an important role. 
It relates the unitary matrices $U_L$ and $U_R$ to each other since the 
triplet induced Majorana mass matrices in Eq.\ (\ref{eq:mogem}) 
have the same coupling matrix $f$ in generation space: 
\be \label{eq:mlmr}
\bad
m_L = f \, v_L & \mbox{ and } & M_R = f \, v_R   \; . 
\ea
\ee
The numbers $v_{L, R}$ are the vacuum expectation values 
(vevs) of the left-- and right--handed 
Higgs triplets, whose existence is needed to maintain the left--right 
symmetry. They receive their vevs at the minimum of the potential, 
producing at the same time masses for the gauge bosons. 
In general \cite{LR}, this results in 
\be \label{eq:vlvr}
v_L \, v_R \simeq \gamma \, M_W^2 , 
\ee
where the constant $\gamma$ is a model dependent parameter of 
$\cal{O}$(1). Inserting this equation as well as Eq.\ (\ref{eq:mlmr}) in 
(\ref{eq:mnu}) yields
\be \label{eq:mnulr}
m_\nu = v_L \, 
\left( f - \tilde{m}_D \, 
\frac{f^{-1}}{\gamma M_W^2} \, \tilde{m}_D^T \right) . 
\ee
\\
As stated before, the contribution of the left-- handed Higgs triplet 
to $m_\nu$ is often neglected. 
However, if one 
compares the relative magnitude of the two contributions 
in Eq.\ (\ref{eq:mnu}), 
denoting the largest mass in the Dirac matrix with $m$, 
one finds 
\be \label{eq:estlr}
\frac{\D |\tilde{m}_D \, M_R^{-1} \, \tilde{m}_D^T |}{\D |m_L| } \simeq 
\frac{m^2/v_R}{\D v_L} \simeq  
\frac{\D m^2}{\D \gamma \, M_W^2}  
\ee
Here, we only used Eq.\ (\ref{eq:vlvr}) and assumed that 
the matrix elements of $f$ and $f^{-1}$ are of the same order of 
magnitude. It is seen that this ratio 
is of order one only for the top quark mass, i.e.\ if one identifies 
the Dirac mass matrix with the up quark mass matrix. Due to the 
hierarchical structure of quark masses, the up quark mass matrix 
can be written as 
\be \label{eq:tmd}
\tilde{m}_D \simeq {\rm diag} (0,0,m_t)   , 
\ee
where $m_t$ denotes the top quark mass. 
This means that only the $(33)$ entry of $m_\nu$ has a contribution 
from $\tilde{m}_D \, M_R^{-1} \, \tilde{m}_D^T $. One can then solve 
for this term and with the experimental knowledge of $m_\nu$ obtain 
$f$. With $f$ and Eq.\ (\ref{eq:mlmr}) we have $M_R$, whose diagonalization 
gives $U_R$, which in turn gives $m_D$ through  
Eqs.\ (\ref{eq:mdur},\ref{eq:tmd}). Then, via Eq.\ (\ref{eq:eps}), one 
obtains the lepton asymmetry. 
This program has been performed in 
\cite{eapjos}, finding that from the solutions to the solar neutrino 
problem only the small angle MSW and vacuum oscillations give an 
asymmetry consistent with the experimental bound (\ref{eq:YBexp}). 
The large angle MSW solution gives a very large 
asymmetry and must be suppressed by fine--tuning the phases 
in the mixing matrix. 
The identification of $\tilde{m}_D$ with the up quark mass matrix 
follows in simple $SO(10)$ models with a 10--plet of Higgses 
generating the fermion masses. The mass relation is different in more 
general situations, when the Higgses belong to the 126 representation 
\cite{moha1}. There are also examples where the Dirac masses for neutrinos 
are zero at tree level and are generated by radiative corrections 
\cite{moha2}. 
In these cases the scale of the Dirac mass is much smaller.\\
 
If we now identify $\tilde{m}_D$ with the down quark or charged lepton mass 
matrix, then the ratio in Eq.\ (\ref{eq:estlr}) is always much smaller than 
one, so that the second term in Eq.\ (\ref{eq:mnulr}) can be 
neglected and it follows
\be \label{eq:fismnu}
f \simeq \frac{1}{v_L} \, m_\nu . 
\ee
Therefore, with the help of 
Eqs.\ (\ref{eq:uldef},\ref{eq:urdef},\ref{eq:mlmr}), one arrives at   
a very simple connection between the left-- and right--handed 
neutrino sectors:  
\be \label{eq:cons}
U_R = U_L \mbox{ and } M_i = m_i \, \frac{v_R}{v_L} 
\ee
The striking property is that the light neutrino masses are 
proportional to the heavy ones. In addition, the left-- and right--handed 
rotation is the same. 
This is the main result of this work and will be used in the 
following analysis\footnote{In the language of \cite{eapjos} one can say 
that the work in this paper represents the special case 
$s=0$ and $m_t \ra m$.}.\\
 
We finally specify the order of 
magnitude of $v_{L,R}$. The maximal size of 
$m_\nu=v_L \, f$ is $\sqrt{\dma}\ls 0.1$ eV, which is 
only compatible with  $v_L\, v_R \simeq \gamma \, M_W^2$ for $v_L \simeq 0.1$ 
eV and $v_R \simeq 10^{15}$ GeV,  
as long as $f \simeq 0.1 \ldots 1$. 
This means that $v_R$ is close to the 
grand unification scale and $v_L$ is of order of the neutrino masses, 
which is expected since $m_L$ is the only contribution to 
$m_\nu$.

\section{\label{sec:estYB}Estimates}

\subsection{\label{sec:mnu}Neutrino mass matrix}
The matrix $U_L$ is measured in neutrino oscillation experiments, 
a convenient parametrisation is 
\bea \label{eq:Upara}
U_L = U_{\rm CKM} \; 
{\rm diag}(1, e^{i \alpha}, e^{i (\beta + \delta)}) \\[0.3cm]
= \left( \bad 
c_1 c_3 & s_1 c_3 & s_3 e^{-i \delta} \\[0.2cm] 
-s_1 c_2 - c_1 s_2 s_3 e^{i \delta} 
& c_1 c_2 - s_1 s_2 s_3 e^{i \delta} 
& s_2 c_3 \\[0.2cm] 
s_1 s_2 - c_1 c_2 s_3 e^{i \delta} & 
- c_1 s_2 - s_1 c_2 s_3 e^{i \delta} 
& c_2 c_3\\ 
               \ea   \right) 
 {\rm diag}(1, e^{i \alpha}, e^{i (\beta + \delta)}) , 
\eea
where $c_i = \cos\theta_i$ and $s_i = \sin\theta_i$. This matrix connects 
gauge eigenstates $\nu_\alpha$ ($\alpha = e, \mu, \tau$) with 
mass eigenstates $\nu_i$ ($i=1,2,3$), i.e.\ 
\[
\nu_\alpha = U_{L\alpha i} \, \nu_i . 
\]  
Note that we identify the 
neutrino mixing matrix 
in Eq.\ (\ref{eq:Upara}) with the matrix $U_L$ diagonalizing the neutrino
mass matrix Eq.\ (\ref{eq:uldef}). 
This assumes implicitly that the charged lepton mixing
is small.\\

The ``CKM--phase'' 
$\delta$ may be probed in oscillation experiments, as long as 
the large mixing angle solution is the solution 
to solar oscillations \cite{osccp}. 
The other 
two ``Majorana phases'' $\alpha$ and $\beta$ can be investigated in 
\onbb{} \cite{Majpha,ichNPB}.  
The choice of the 
parameterization in Eq.\ (\ref{eq:Upara}) reflects this fact since 
the $ee$ element of the mass matrix $\sum_i U_{Lei}^2 m_i$ 
depends only on the phases 
$\alpha$ and $\beta$. In a hierarchical scheme, to which we will 
limit ourselves, there is no constraint 
on the phases from \onbb{} \cite{ichNPB}. 
Thus, we can choose them arbitrarily. The mass eigenstates are given as 
\be \label{eq:m3m2m1}
\ba
m_3 = \sqrt{\dma + m_2^2} \simeq \sqrt{\dma} \\[0.4cm]
m_2 = \sqrt{\dms + m_1^2} \simeq \sqrt{\dms} \gg m_1  
\ea . 
\ee
The values of $\theta_2$ and \dma{} are known to a good 
precision, corresponding to maximal mixing $\theta_2 \simeq \pi/4$ and 
$\dma \simeq 3 \times 10^{-3}$ eV$^2$. Regarding $\theta_1$ and \dms{} 
three distinct areas in the parameter space are allowed, small (large) 
mixing, denoted by SMA (LMA) and quasi--vacuum oscillations (QVO): 
\be \label{eq:solsol}
\bad
\mbox{ SMA: } & \tan^2 \theta_1 \simeq 10^{-4} \ldots 10^{-3} , & 
\dms \simeq 10^{-6} \ldots 10^{-5} \, {\rm eV^2} \\[0.4cm]
\mbox{ LMA: } & \tan^2 \theta_1 \simeq 0.1 \ldots 4 , & 
\dms \simeq 10^{-5} \ldots 10^{-3} \, {\rm eV^2} \\[0.4cm]
\mbox{ QVO: } & \tan^2 \theta_1 \simeq 0.2 \ldots 4 , & 
\dms \simeq 10^{-10} \ldots 10^{-7} \, {\rm eV^2}    
\ea  . 
\ee
For the last angle $\theta_3$ there exists only a limit of about 
$\sin^2 \theta_3 \ls 0.08$. See \cite{carlos} for a recent 
three--flavor fit to all available data and \cite{mnu} for 
a more general discussion of the derivation of neutrino mass matrices.\\
 
The results for atmospheric and solar mixing imply two simple forms 
for $U_L$, namely 
\bea \label{eq:ullm}
U_L \simeq \left( \bad 
\frac{1}{\sqrt{2}} & \frac{1}{\sqrt{2}} & s_3 e^{-i \delta} \\[0.2cm] 
-\frac{1}{2} (1 + s_3 e^{i \delta})  
& \frac{1}{2} (1 - s_3 e^{i \delta})  
& \frac{1}{\sqrt{2}} \\[0.2cm] 
\frac{1}{2} (1 - s_3 e^{i \delta})  
& -\frac{1}{2} (1 + s_3 e^{i \delta})  
& \frac{1}{\sqrt{2}}\\ 
               \ea   \right) 
 {\rm diag}(1, e^{i \alpha}, e^{i (\beta + \delta)})  
\eea 
for LMA as well as QVO and 
\bea \label{eq:ulsm}
U_L \simeq \left( \bad 
1 & 0 & s_3 e^{-i \delta} \\[0.2cm] 
-\frac{1}{2} s_3 e^{i \delta}  
& \frac{1}{\sqrt{2}}   
& \frac{1}{\sqrt{2}} \\[0.2cm] 
-\frac{1}{2} s_3 e^{i \delta}  
& -\frac{1}{\sqrt{2}}   
& \frac{1}{\sqrt{2}}\\ 
               \ea   \right) 
 {\rm diag}(1, e^{i \alpha}, e^{i (\beta + \delta)})  
\eea 
for the SMA case. 
These forms can be used to estimate the magnitude of the 
baryon asymmetry within our scenario.

\subsection{\label{sec:ooe}Out--of--equilibrium condition}
Besides $CP$ and lepton--number violation, a necessary condition 
\cite{sakharov} for leptogenesis to work 
is the out--of--equilibrium decay of the 
heavy Majorana neutrinos, i.e.\ 
\be \label{eq:K}
K \equiv \frac{\Gamma_1}{H(T = M_1)} = 
\frac{ (m_D^\dagger m_D)_{11} \, M_1}{8\, \pi \, v^2}  
\frac{M_{\rm Pl}}{ 1.66 \, \sqrt{g^\ast} \, M_1^2} \le 1 , 
\ee 
where $\Gamma_1$ is the width of the lightest Majorana neutrino and 
$H(T = M_1)$ the Hubble constant at the temperature of the decay. 
The number of massless degrees of freedom at the time of the decay is 
$g^\ast \simeq 110$ and 
$M_{\rm Pl}$ is the Planck mass. However, values of $K$ smaller than ten 
still give sizable lepton asymmetry, even though the asymmetry is reduced by 
lepton--number violating wash--out processes. This suppression is 
obtained by integrating the Boltzmann equations and can be 
parameterized by \cite{param}
\be \label{eq:kappa}
\kappa \simeq 
\left\{ 
\bad 
\sqrt{0.1 \, K} \, \exp (-4/3 \, (0.1\, K)^{0.25}) 
& \mbox{ for } & K \gs 10^6 \\[0.4cm]
\frac{\D 0.3}{\D K \, (\ln K)^{0.6} }
&  \mbox{ for } & 10 \ls K \ls 10^6 \\[0.4cm]
\frac{\D 1}{\D 2 \, \sqrt{K^2 \, + 9}} 
& \mbox{ for } & 0 \ls K \ls 10
\ea 
\right. . 
\ee
Since also the down quarks and charged leptons display a mass hierarchy, 
we may write the Dirac mass matrix in analogy to Eq.\ (\ref{eq:tmd}) as 
\be \label{eq:tmd1}
\tilde{m}_D \simeq {\rm diag} (0, 0, m) , 
\ee
where $m$ might be the bottom quark or tau lepton mass. 
Inserting this in Eq.\ (\ref{eq:mdur}) and using the approximate forms for 
$U_L$ in Eqs.\ (\ref{eq:ullm},\ref{eq:ulsm}) yields 
\be \label{eq:md11}
(m_D^\dagger m_D)_{11} = |U_{L\tau 1}|^2 \, m^2 \simeq \frac{m^2}{4}
\left\{ \baz  (1 - 2 \, s_3 \, c_\delta ) & \mbox{ for LMA, QVO}\\[0.3cm]
             s_3^2 & \mbox{ for SMA} \ea \right.  , 
\ee
where we took only the leading term in $s_3$ and set $c_3=1$. 
Using Eq.\ (\ref{eq:cons}) we replace $M_1$ by $\frac{v_R}{v_L} \, m_1$ 
and all other quantities by known constants to arrive at 
\be \label{eq:estK}
K \simeq 1.5 \, \left(\frac{10^{-6} \, \rm eV}{m_1} \right) 
\left(\frac{10^{15} \, \rm GeV}{v_R} \right)^2 \gamma  
\left(\frac{m}{\rm GeV} \right)^2 
\left\{ \baz (1 - 2 \, s_3 \,  c_\delta ) & \mbox{ for LMA, QVO}\\[0.3cm]
             s_3^2 & \mbox{ for SMA} \ea \right.  .  
\ee 
Thus, for typical 
parameters, the dilution factor is not too 
small. We shall assume for the following estimates the value 
$\kappa \ls 1/6$.

\subsection{\label{sec:eps}Lepton and baryon asymmetry}
From Eqs.\ (\ref{eq:mdur},\ref{eq:tmd1}) and the approximate forms of 
$U_L$ we calculate the imaginary parts of $(m_D^\dagger m_D)_{1j}$ 
needed for calculating the lepton asymmetry in Eq.\ (\ref{eq:eps}). 
For the 12 element  
\bea \label{eq:immd1112} 
{\rm Im} (m_D^\dagger m_D)^2_{12} = 
m^4 \, {\rm Im} (U_{L\tau 1}^\ast U_{L \tau 2})^2 \\[0.5cm]
\simeq \frac{\D m^4}{\D 16} 
\left\{ \baz 
 (s_{2\alpha} + 4 \, s_3 \, s_{\delta}\, c_{2\alpha}) 
& \mbox{ for LMA, QVO}\\[0.3cm]
            2 \, s_3^2 \, s_{2(\alpha-\delta)} & \mbox{ for SMA} 
\ea \right. 
\eea 
and for the 13 element 
\bea \label{eq:immd1113} 
{\rm Im} (m_D^\dagger m_D)^2_{13} = 
m^4 \, {\rm Im} (U_{L\tau 1}^\ast U_{L \tau 3})^2 \\[0.5cm]
\simeq \frac{\D m^4}{\D 8} 
\left\{ \baz 
(s_{2(\beta + \delta)} - 2 \, s_3 \, s_{2 \beta+\delta})     
& \mbox{ for LMA, QVO}\\[0.3cm]
          s_3^2  \, s_{2 \beta}  & \mbox{ for SMA} 
\ea \right. 
\eea
with the notation 
$s_{2(\alpha-\delta)} = \sin 2 (\alpha - \delta)$ and so on. 
Note that the asymmetry $\varepsilon$ is proportional to 
$(m_D^\dagger m_D)^{-1}_{11}$ and therefore to $s_3^{-2}$ for the SMA case. 
This might enhance the lepton asymmetry dramatically. 
However, as seen from the last two equations, for the SMA case,  
${\rm Im} (m_D^\dagger m_D)^2_{1j}$ is proportional to $s_3^2$, which 
cancels this potentially dangerous term.\\

From the lepton asymmetry $\varepsilon$ the baryon asymmetry $Y_B$ 
is obtained by 
\be \label{eq:YBth}
Y_B = c\, \kappa \frac{\varepsilon}{g^\ast} , 
\ee
where $c \simeq - 0.55$, denoting the fraction of the lepton asymmetry 
converted into a baryon asymmetry via sphaleron \cite{sphaleron} 
processes.   
Using the approximate 
form of the function $f$ in Eq.\ (\ref{eq:fapprox}), the mass 
eigenstates from Eq.\ (\ref{eq:m3m2m1}) and the fact that 
$M_i/M_j = m_i/m_j$ one finally finds for the baryon asymmetry for 
the LMA and QVO case 
\be \label{eq:estYBLM}
\bad Y_B \cdot 10^{10} \ls 
4.1 \, \left( \frac{\D m}{\D \rm GeV} \right)^2 
\frac{\D 1}{\D 1 - 2 \, s_3 \, c_\delta } 
\left\{
(s_{2\alpha} + 4 \, s_3 \, s_{\delta} \, c_{2\alpha}   )
\frac{\D m_1}{\D \sqrt{\dms}} \right. \\[0.5cm] 
\left. + 2 (s_{2(\beta + \delta) }- 2 \, s_3\, s_{2\beta+\delta}) 
\frac{\D m_1}{\D \sqrt{\dma}} 
\right\}  \ea 
\ee
whereas in the SMA case one gets
\be \label{eq:estYBSM}
Y_B \cdot 10^{10} \ls 
8.2 \, \left( \frac{\D m}{\D \rm GeV} \right)^2  
\left\{ s_{2 (\alpha - \delta)} \frac{\D m_1}{\D \sqrt{\dms}} + 
s_{2 \beta} \frac{\D m_1}{\D \sqrt{\dma}}
\right\} . 
\ee
Thus, for comparable values of the phases and $\dms \ll \dma$ the first 
term dominates $Y_B$. Note that $Y_B$ vanishes for 
$\alpha = \beta = \delta = 0$, in which $CP$ is conserved, and that it is  
proportional to the square of the largest mass in the Dirac mass matrix.   
Therefore, if one would take a charged lepton mass matrix, the 
results for $Y_B$ will change roughly by a factor 
of $\frac{m_\tau^2}{m_b^2} \simeq 0.2$.\\

To sum up, identifying the Dirac mass matrix with the down quark 
or charged lepton mass matrix we obtain the 
simple relations in Eq.\ (\ref{eq:cons}) which introduces the 
lightest left--handed neutrino mass in the answers. In our 
previous paper \cite{eapjos} the presence of the top quark mass did 
not permit such a simplification because a different 
diagonalization for $U_L$ and $U_R$ had to be performed. As a 
consequence, $M_1$ was not directly proportional to $m_1$ but 
received contributions from $m_2$, $m_3$ and a term of order 
$m_t\, M_W/v_R$. These contributions are all larger than $m_1$ 
so that basically no dependence on $m_1$ was evident.

\subsection{\label{sec:limm1}Limit on smallest mass eigenstate}
The approximate form of $Y_B$ in 
Eqs.\ (\ref{eq:estYBLM},\ref{eq:estYBSM}) can be used 
to derive a limit on the smallest mass eigenstate $m_1$. In principle, 
it can be exactly zero, since oscillation experiments only measure 
the difference of the square of two masses. 
As mentioned already, in \cite{eapjos} there was practically 
no dependence on $m_1$ whereas here it is decisive since $Y_B$ is 
directly proportional to $m_1$. 
To get the highest possible $Y_B$ we choose 
the phases such that their contribution is maximal and take the 
lowest possible \dms, as listed in 
Eq.\ (\ref{eq:solsol}). The factor $(1 - 2 \, s_3 \, c_\delta )^{-1}$ 
can give for $s_3 \ls 0.3$ give 
an enhancement of about 2.5 at most 
so that for $m=m_b$ we obtain:
\be \label{eq:YBmax}
Y_B \cdot 10^{10} \ls \left\{
\bav 
\left( \frac{\D m_1}{10^{-5}\, \rm eV} \right) & \cdot & 
0.6 & \mbox{ for SMA} \\[0.3cm]
\left( \frac{\D m_1}{10^{-6}\, \rm eV} \right) & \cdot & 
1.4 & \mbox{ for LMA} \\[0.3cm]
\left( \frac{\D m_1}{10^{-7}\, \rm eV} \right) & \cdot & 
2.1 & \mbox{ for QVO} \\[0.3cm]  
\ea \right. 
\ee   
Therefore, to obtain an asymmetry within the experimental range, 
$m_1$ should not be much smaller than $10^{-6}$ ($10^{-7} ,\, 10^{-8}$) 
eV for the SMA (LMA, QVO) case.

\section{\label{sec:res}Numerical Results}
\begin{figure}[hp]
\epsfig{file=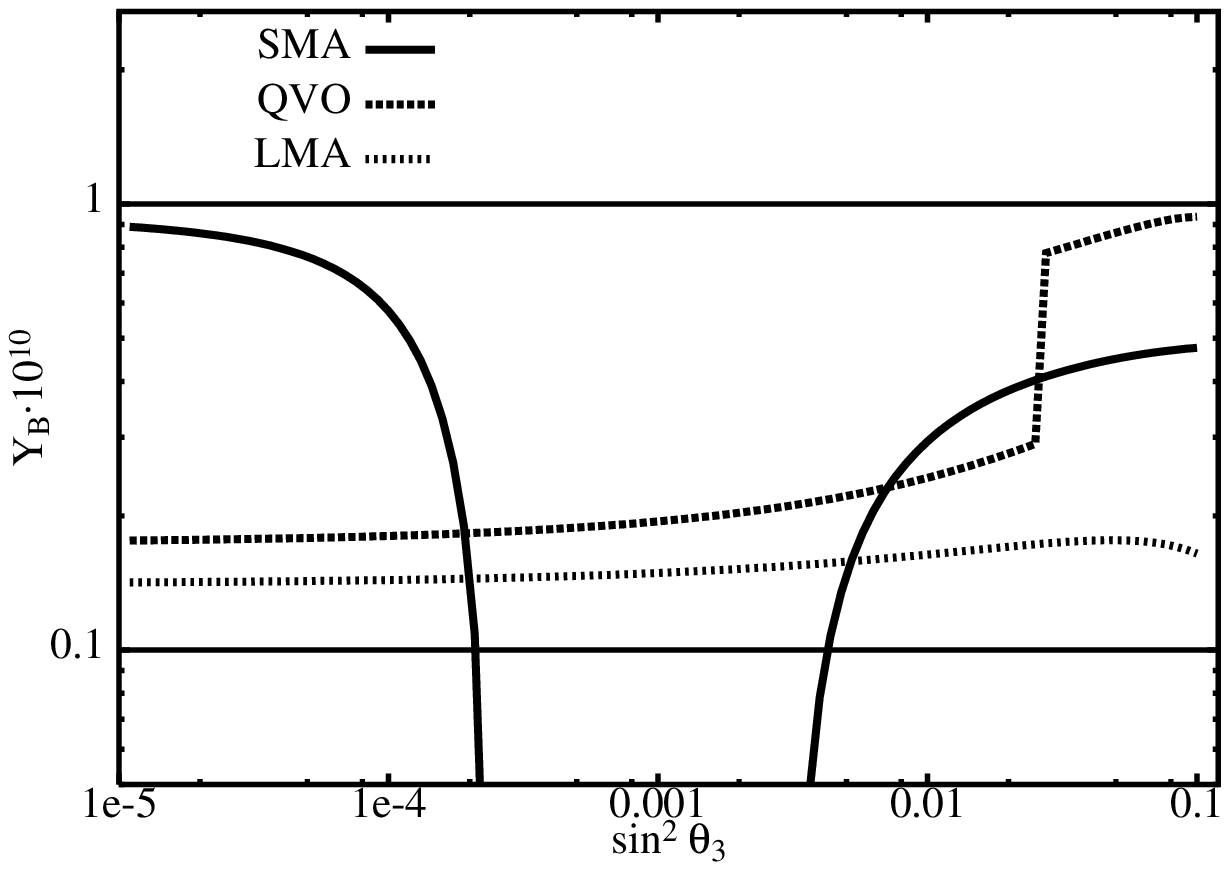,width=13cm,height=8cm}
%\vspace{0.5cm}
\caption{\label{fig:YB}Behavior of the baryon asymmetry as a function of 
$\sin^2 \theta_3$ for the 
parameters as given in Sec.\ \ref{sec:res}. 
For this plot, the Dirac mass matrix is a down quark mass matrix.}
\vspace{0.5cm}
\epsfig{file=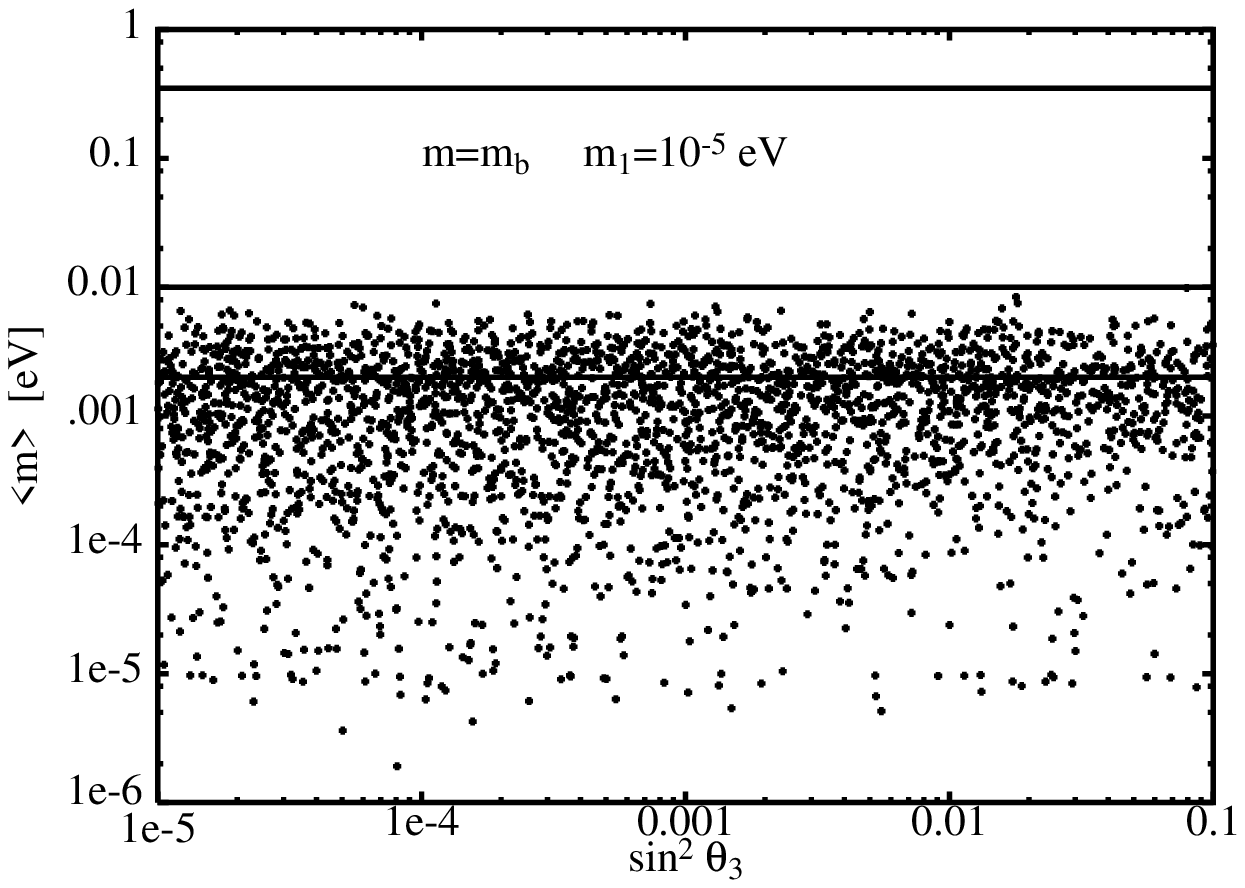,width=13cm,height=8cm}
%\vspace{0.5cm}
\caption{\label{fig:meemb5}The effective electron neutrino mass \meff 
as a function of $\sin^2 \theta_3$ for $m_1 = 10^{-5}$ eV\@. 10000  
random points were generated, about 2700 give $Y_B$ within 0.1 and 1 and 
less than 800 an effective mass \meff above 0.002 eV\@.  
For this plot, the Dirac mass matrix is a down quark mass matrix.}
\end{figure}

To check the estimates made in the previous section we show in 
Fig.\ \ref{fig:YB} the baryon asymmetry for three typical values of the 
three solar solutions, 
\be \label{eq:typ}
\bad
\dms = 5 \cdot 10^{-6} \, {\rm eV^2},  
& \tan^2 \theta_1 = 5 \cdot 10^{-4} & \mbox{ SMA }\\[0.4cm]
\dms = 5 \cdot 10^{-5} \, {\rm eV^2},  
& \tan^2 \theta_1 = 1 & \mbox{ LMA }\\[0.4cm]
\dms =  10^{-8} \, {\rm eV^2}, 
& \tan^2 \theta_1 = 1 & \mbox{ QVO }
\ea , 
\ee
fixed $\dma = 3 \cdot 10^{-3}$ eV$^2$, $\theta_2 = \pi/4$ and choose 
the phases 
$3 \, \alpha = 5 \, \beta = 6 \, \delta = \pi$. The two parameters 
of the left--right symmetry, $v_R$ and $\gamma$ were taken 
$10^{15}$ GeV and $1$, respectively. 
For $\tilde{m}_D$ we took 
a typical down quark mass matrix (see e.g.\ \cite{models})
\be \label{eq:tmdtyp}
\tilde{m}_D = \left( \bad 
0                 & \sqrt{m_d \, m_s} & 0 \\[0.3cm]
\sqrt{m_d \, m_s} & m_s               &  \sqrt{m_b \, m_s} \\[0.3cm]
0                 & \sqrt{m_b \, m_s} & m_b 
\ea \right) ,   
\ee
where $m_{d,s,b}$ are the masses of the $d$, $s$ and $b$ quark. 
The lowest mass eigenstate $m_1$ is chosen $10^{-5}$ $(2 \cdot 10^{-6})$
eV for the SMA and LMA (QVO) case.  
We find that there is practically no difference when one takes 
this matrix or the simple form in Eq.\ (\ref{eq:tmd1}), showing 
the model independence of our scenario.\\
 
For the SMA case there are areas, in which $Y_B$ becomes very small 
and even negative. They are not found by our 
estimates and can be shown to stem from cancellations of terms 
proportional to $\sin^2 \theta_3$ and $\sin^2 \theta_1$. There is 
no significant dependence on $\sin^2 \theta_3$ for the LMA 
case, as also indicated 
by the approximate expressions in 
Eqs.\ (\ref{eq:estYBLM},\ref{eq:estYBSM}). Regarding QVO, one sees that 
for larger $s_3$ the dilution factor rises from about 0.01 to 
its maximal value. This explains the rise of $Y_B$ with $s_3^2$ for 
this solution.  
With comparable phases, the asymmetry of the LMA and SMA case can be 
of the same order whereas the one for QVO needs suppression through 
$m_1$ due to the very low \dms. If one inserts our chosen parameters 
in the approximate forms for $K$ and $Y_B$ in 
Eqs.\ (\ref{eq:estK},\ref{eq:estYBLM},\ref{eq:estYBSM}), then one 
finds for $s_3 = 0$ that 
\be \label{eq:YBapprox}
Y_B \cdot 10^{10} \simeq \left\{ \baz 
0.7 & \mbox{ for SMA} \\[0.3cm]
0.1 & \mbox{ for LMA} \\[0.3cm] 
0.1 & \mbox{ for QVO}  
\ea \right. 
\ee
which is in good agreement with the exact result.

\section{\label{sec:0vbb}Connection to Neutrinoless Double Beta Decay}
As mentioned before, only the LMA solution provides the possibility 
to find leptonic $CP$ violation in oscillation experiments \cite{osccp}.  
In addition, in the hierarchical mass scheme only the LMA solution 
might produce a measurable Majorana mass for the electron 
neutrino \cite{bilpet}. Fortunately, the 
latest SuperKamiokande data \cite{SK} seems to indicate 
that LMA is the preferred 
solution of the solar neutrino problem.  
The Majorana mass of the electron neutrino is defined as  
\be \label{eq:meff}
\meff \! = \sum_i \, U_{L ei}^2 \, m_i
\ee
and due to the complex matrix elements $U_{L \alpha i}$ there is a 
possibility of cancellation \cite{ichNPB} of terms in Eq.\ (\ref{eq:meff}). 
It is therefore interesting to ask if the parameters 
that produce a satisfying $Y_B$ also deliver a sizable 
$\meff\!\!$. In Fig.\ \ref{fig:meemb5} we show the 
results of a random scan of the parameter space 
with a down quark Dirac 
mass matrix, taking the following values: \dms{} between 
$10^{-5}$ and $10^{-3}$ eV$^2$, $\tan^2 \theta_1$ between 0.1 and 4 and 
the three phases between zero and $2 \pi$. The atmospheric parameters 
were fixed as before.  
We choose for the lowest mass eigenstate $m_1= 10^{-5}$ eV\@. 
Included in the plot is the current limit of $\meff \! \! \!\ls 0.35$ eV 
\cite{mefflim} and 
two future limits, 0.01 and 0.002 eV\@. From 10000 parameter sets, 
about 2700 give a baryon asymmetry of the correct magnitude. However, 
most of the points (about 2000) lie below the 
lowest achievable \meff limit 
of 0.002 eV\@. For higher $m_1$ 
the fraction of parameter sets giving an acceptable $Y_B$ decreases. 
For $m_1 = 10^{-3}$ eV, only 5 $\%$ give acceptable $Y_B$ 
and \meff$\!\!$.\\
 
\begin{figure}[ht]
\epsfig{file=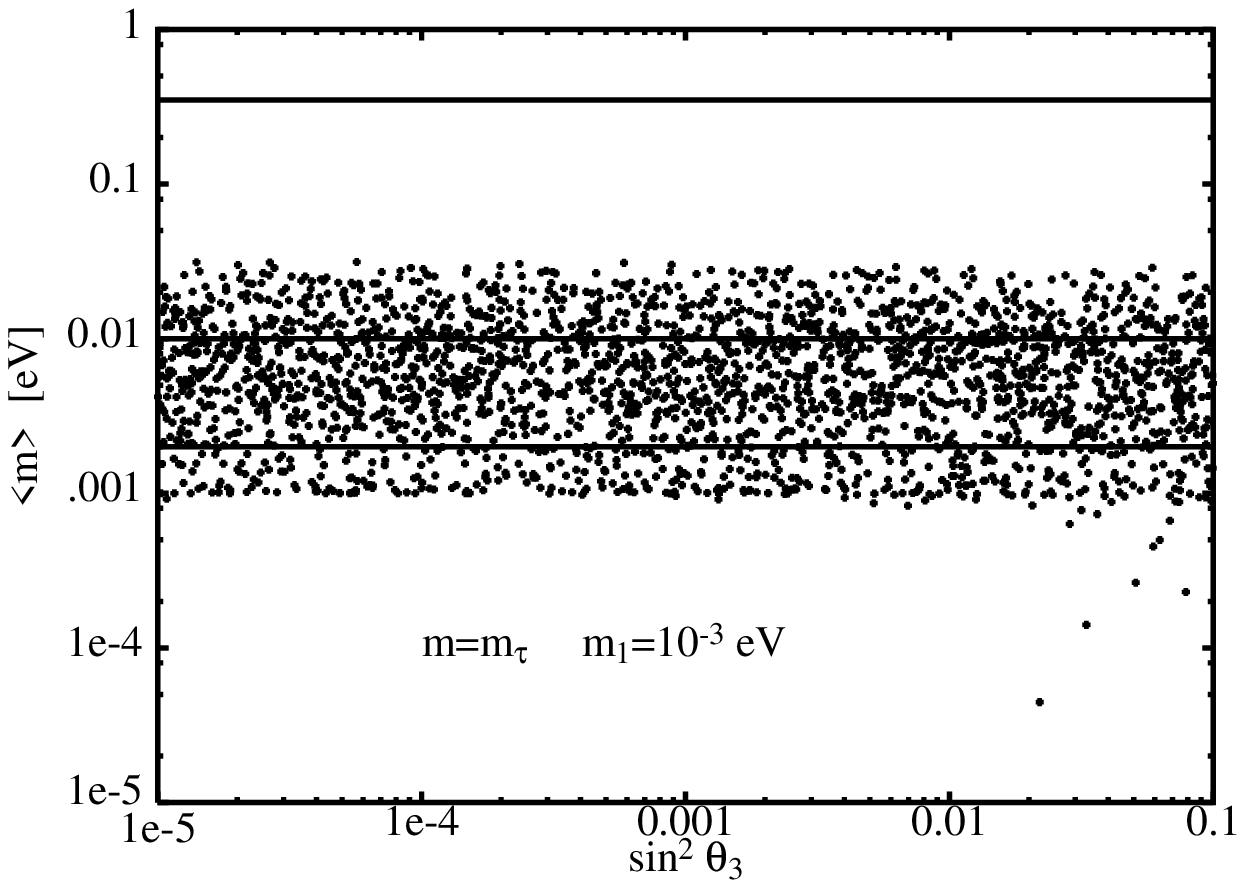,width=13cm,height=8cm}
%\vspace{0.5cm}
\caption{\label{fig:meemt3}The effective electron neutrino mass \meff 
as a function of $\sin^2 \theta_3$ for $m_1 = 10^{-3}$ eV\@. 7000  
random points were generated, about 2500 give $Y_B$ within 0.1 and 1 and 
about 2100 an effective mass \meff above 0.002 eV\@. 
For this plot, the Dirac mass matrix is a charged lepton mass matrix.}
\end{figure}

If on the other hand we identify the Dirac mass matrix with a 
charged lepton mass matrix, the asymmetry is reduced. A charged lepton mass 
matrix is obtained from 
Eq.\ (\ref{eq:tmdtyp}) by replacing the bottom (charm, down) quark 
mass with the tau (muon, electron) mass.  
Then, the asymmetry needs a smaller suppression through the phases and 
\meff can be bigger, going up to its maximal value of about 
$\frac{1}{2} \, \sqrt{\dms_{\rm max}} \simeq 0.016$ eV\@.  
This can be seen in Fig.\ \ref{fig:meemt3} for $m_1 = 10^{-3}$ eV\@. 
Here, about 35 $\%$ of the parameter sets give a correct $Y_B$ and 
29 $\%$ an \meff above 0.002 eV\@. For lower 
$m_1$, this fraction decreases rapidly because the asymmetry 
decreases. 

The ``CKM--phase'' $\delta$ 
does not appear in \meff$\!\!$. It appears in the asymmetry but 
turns out to be not very strongly restricted. The reason is that in our  
parametrisation (\ref{eq:Upara}) the phase $\delta$ 
is often connected with the small quantity $s_3$, which results 
in a small overall dependence of the asymmetry on 
this phase. 
Therefore, one can expect a sizable  
\meff within our scenario as long as the smallest mass eigenstate 
$m_1$ is not much lighter than $10^{-3}$ eV and 
the Dirac mass matrix is identified with the charged lepton 
mass matrix.   
The possibility of detecting 
$CP$ violating effects in oscillation experiments remains open, 
in the sense that within our scenario the phase $\delta$ can take any value.  
Finally, we remark that for the scenario in \cite{eapjos} an extreme 
fine--tuning of the parameters is required to obtain a reasonable $Y_B$. 
Therefore, no sizable signal in \onbb{} can be expected if one 
identifies the Dirac mass matrix with the up quark mass matrix.

%\newpage

\section{\label{sec:concl}Conclusions}
In a left--right symmetric theory with a left--handed Higgs triplet 
the light neutrino mass matrix originates from its vev. 
If in addition the Dirac mass matrix in the see--saw formula 
is identified with the down quark or charged lepton mass matrix, 
then a simple connection between the light left--handed 
and heavy right--handed neutrino sectors emerges. 
The out--of--equilibrium 
condition for the decay of the heavy Majorana neutrinos is 
automatically fulfilled and a baryon asymmetry of the correct 
order of magnitude is produced. 
The limit on the smallest mass eigenstate $m_1$ is approximately  
$10^{-6}$ ($10^{-7} ,\, 10^{-8}$) for the SMA (LMA, QVO) cases. 
Within this scenario a sizable effect in neutrinoless double beta 
decay is expected, as long as $m_1$ is not lighter than 
$10^{-3}$ eV and 
the Dirac mass matrix is the charged lepton mass matrix. 
The possibility of finding 
$CP$ violation in oscillation experiments remains open.

\hspace{3cm}
\begin{center}
{\bf \large Acknowledgments}
\end{center}
This work has been supported in part by the
``Bundesministerium f\"ur Bildung, Wissenschaft, Forschung und
Technologie'', Bonn under contract No. 05HT1PEA9.
Financial support from the Graduate College
``Erzeugung und Zerf$\ddot{\rm a}$lle von Elementarteilchen''
at Dortmund university is gratefully acknowledged (W.R.). W.R.\ wishes 
to thank the Universita degli studi di Pisa 
where part of this work was performed. A.S.J.\ acknowledges
support from the Alexander von Humboldt foundation and 
thanks his colleagues for their 
hospitality at the University of Dortmund.

\begin{thebibliography}{99}
\bibitem{nurev}K. Zuber, \Jo{\PRP}{305}{295}{1998}; 
S. M. Bilenky, C. Giunti, and W. Grimus, 
\Jo{\PNPP}{43}{1}{1999}; 
J. Ellis, Talk given at the 19th International Conference on Neutrino 
Physics and Astrophysics - Neutrino 2000, Sudbury, Ontario, Canada, 
16-21 Jun 2000, \Jo{\NPBP}{91}{503}{2000}; 
P. Langacker, Summary talk for Europhysics Neutrino Oscillation 
Workshop (NOW 2000), Conca Specchiulla, Otranto, Lecce, Italy, 
9-16 Sep 2000, \Jo{\NPBP}{100}{383}{2001}.
%%CITATION = PRPLC,305,295;%%
%%CITATION = PPNPD,43,1;%%
%%CITATION = NUPHZ,91,503;%%
%%CITATION = NUPHZ,100,383;%%
\bibitem{SMnot}G. R. Farrar, M. E. Shaposhnikov, \Jo{\PRD}{50}{774}{1994}.
%%CITATION = PHRVA,D50,774;%%
\bibitem{first}M. Fukugita, T. Yanagida, \Jo{\PLB}{174}{45}{1986}.
\bibitem{lola}J. Ellis, S. Lola, and D. V. Nanopoulos,
\Jo{\PLB}{452}{87}{1999}.
%%CITATION = HEP-PH 9902364;%%
\bibitem{laza1}G. Lazarides, N. D. Vlachos, \Jo{\PLB}{459}{482}{1999}.
%%CITATION = HEP-PH 9903511;%%   
\bibitem{others}M. S. Berger and B. Brahmachari, \Jo{\PRD}{60}{073009}{1999}; 
M. S. Berger, \Jo{\PRD}{62}{013007}{2000}.
W. Buchm$\rm \ddot{u}$ller and M. Pl$\rm \ddot{u}$macher, 
hep-ph/0007176. 
%%CITATION = PHRVA,D60,073009;%%
%%CITATION = PHRVA,D62,013007;%%
%%CITATION = HEP-PH 0007176;%% 
\bibitem{kang}
K. Kang, S. K. Kang, and U. Sarkar, \Jo{\PLB}{486}{391}{2000}.  
%%CITATION = PHLTA,B486,391;%%
\bibitem{goldberg}
H. Goldberg, \Jo{\PLB}{474}{389}{2000}. 
%%CITATION = PHLTA,B474,389;%%
\bibitem{nezri}E. Nezri, J. Orloff, hep-ph/0004227. 
%%CITATION = HEP-PH 0004227;%%
\bibitem{faltra1}
D. Falcone, F. Tramontano, \Jo{\PRD}{63}{073007}{2001}.   
%%CITATION = PHRVA,D63,073007;%%
\bibitem{faltra2}
D. Falcone, F. Tramontano, \Jo{\PLB}{506}{1}{2001}. 
%%CITATION = PHLTA,B506,1;%%
\bibitem{nielsen}H. B. Nielsen, Y. Takanishi, hep-ph/0101307. 
%%CITATION = HEP-PH 0101307;%%
\bibitem{eapjos}A. S. Joshipura, E. A. Paschos, hep-ph/9906498;
A. S. Joshipura, E. A. Paschos, and W. Rodejohann, hep-ph/0104228. 
%%CITATION = HEP-PH 9906498;%%
%%CITATION = HEP-PH 0104228;%%
\bibitem{chun}E. Ma, U. Sarkar, \Jo{\PRL}{80}{5716}{1998};
E. Ma, U. Sarkar, and T. Hambye, hep-ph/0011192;
E. J. Chun and S. K. Kang, \Jo{\PRD}{63}{097902}{2001}. 
%%CITATION = PRLTA,80,5716;%% 
%%CITATION = HEP-PH 0011192;%%
%%CITATION = PHRVA,D63,097902;%%
\bibitem{seesaw}M. Gell--Mann and P. Ramond, and R. Slansky, 
in {\it Supergravity}, P. van Nieuwenhuizen $\&$ D. Z. Freedman (eds.), 
North Holland Publ. Co., 1979 p 315; 
T. Yanagida, 
Proc. of the {\it Workshop on Unified Theories and the Baryon
Number of the Universe}, edited by 
O. Sawada and A. Sugamoto, KEK, Japan 1979; 
R. N. Mohapatra, G. Senjanovic, \Jo{\PRL}{44}{912}{1980}.
%%CITATION = PRLTA,44,912;%%      
\bibitem{citeYB}K. A. Olive, G. Steigman, and T. P. Walker, 
\Jo{\PRP}{333}{389}{2000}
%%CITATION = PRPLC,333,389;%%
\bibitem{moh} D. Caldwell and R. N. Mohapatra, \Jo{\PRD}{48}{3259}{1993};
A. S. Joshipura, \Jo{\ZPC}{64}{31}{1994}; B. Brahmachari and
R. N. Mohapatra,
\Jo{\PRD}{58}{015001}{1998}. 
%%CITATION = PHRVA,D48,3259;%%
%%CITATION = ZEPYA,C64,31;%%
%%CITATION = PHRVA,D58,015001;%%  
\bibitem{luty}M. A. Luty, \Jo{\PRD}{45}{455}{1992}.
%%CITATION = PHRVA,D45,455;%%
\bibitem{leptogenesis}M. Flanz, E. A. Paschos, and U. Sarkar,
\Jo{\PLB}{345}{248}{1995};
%%CITATION = PHLTA,B345,248;%%
L. Covi, E. Roulet, and F. Vissani, \Jo{\PLB}{384}{169}{1996};
%%CITATION = PHLTA,B384,169;%%
M. Flanz {\it et al.}, \Jo{\PLB}{389}{693}{1996};
%%CITATION = PHLTA,B389,693;%%
A. Pilaftsis, \Jo{\PRD}{56}{5431}{1997};
%%CITATION = PHRVA,D56,5431;%%
W. Buchm$\rm \ddot{u}$ller and M. Pl$\rm \ddot{u}$macher,
\Jo{\PLB}{431}{354}{1998}.
%%CITATION = PHLTA,B431,354;%%
\bibitem{LR}R. N. Mohapatra and P. B. Pal, 
{\it Massive neutrinos in physics and astrophysics}, 
Singapore, Singapore: World Scientific (1991) 318 p. 
(World Scientific lecture notes in physics, 41).     
\bibitem{moha1}See for example, Brahmachari and Mohapatra in 
Ref.\ \cite{moh}.
\bibitem{moha2}D. Chang and R. N. Mohapatra, \Jo{\PRL}{58}{1600}{1987}; 
R. N. Mohapatra, \Jo{\PLB}{201}{517}{1988}. 
%%CITATION = PRLTA,58,1600;%%
%%CITATION = PHLTA,B201,517;%%
\bibitem{osccp}M. Koike, J. Sato, \Jo{\PRD}{61}{073012}{2000};
%%CITATION = PHRVA,D61,073012;%%
V. Barger {\it et al.}, \Jo{\PRD}{62}{073002}{2000}.
%%CITATION = PHRVA,D62,073002;%%
\bibitem{Majpha}T. Fukuyama {\it et al.}, hep-ph/0012357. 
%%CITATION = HEP-PH 0012357;%%
\bibitem{ichNPB}W. Rodejohann, \Jo{\NPB}{597}{110}{2001}.
%%CITATION = NUPHA,B597,110;%%
\bibitem{carlos}M. C. Gonzalez--Garcia {\it et al.}, 
\Jo{\PRD}{63}{033005}{2001}.
%%CITATION = PHRVA,D63,033005;%%
\bibitem{mnu}
S. M. Bilenky, C. Giunti, \Jo{\PLB}{444}{379}{1998};  
%%CITATION = PHLTA,B444,379;%%
W. Rodejohann, \Jo{\PRD}{62}{013011}{2000}.
%%CITATION = PHRVA,D62,013011;%%
\bibitem{sakharov}A. D. Sakharov, JETP Lett. {\bf 5}, 24 (1967).
%%CITATION = ZFPRA,5,32;%% 
\bibitem{param}E. W. Kolb, M. S. Turner, {\it The early universe}, 
Redwood City, USA: Addison-Wesley (1990), (Frontiers in physics, 69); 
A. Pilaftsis, \Jo{\IJMP}{14}{1811}{1999}; 
E. A. Paschos, M. Flanz, \Jo{\PRD}{58}{113009}{1998}. 
%%CITATION = IMPAE,A14,1811;%%"
%%CITATION = HEP-PH 0101307;%% 
%%CITATION = PHRVA,D58,113009;%%
\bibitem{sphaleron}V. A. Kuzmin, V. A. Rubakov, and M. E. Shaposhnikov, 
\Jo{\PLB}{155}{36}{1985}.
%%CITATION = PHLTA,B155,36;%%
\bibitem{models}B. P. Desai, A. R. Vaucher, \Jo{\PRD}{63}{113001}{2001}.
%%CITATION = PHRVA,D63,113001;%%
\bibitem{bilpet}S. M. Bilenky, S. Pascoli, and S. T. Petcov, hep-ph/0102265.
%%CITATION = HEP-PH 0102265;%%
\bibitem{SK}S. Fukuda {\it et al.}, hep-ex/0103033.
%%CITATION = HEP-EX 0103033;%%
\bibitem{mefflim}H. V. Klapdor--Kleingrothaus {\it et al.}, 
Talk presented at 3rd International Conference on Dark Matter 
in Astro and Particle Physics (Dark2000), Heidelberg, Germany, 
10-16 Jul 2000, hep-ph/0103062.
%%CITATION = HEP-PH 0103062;%%


\end{thebibliography}
\end{document}